\documentclass[aps,floats,prl,nofootinbib]{revtex4}
\usepackage{amsfonts,amsmath,amssymb,ascmac,bm,amsthm}
\usepackage{fnpct} 
\usepackage{comment}
\usepackage{caption}
\usepackage{ifpdf}
\usepackage{graphicx}
\usepackage{slashed}
\usepackage{color}
\usepackage[mathscr]{eucal}
\usepackage[utf8]{inputenc}
\usepackage{physics}
\usepackage{cancel}
\usepackage{float}
\usepackage{soul}
\usepackage{booktabs}
\usepackage{simpler-wick}
\usepackage{hyperref}
\usepackage{tensor}
\usepackage{upgreek}
\usepackage{caption}
\usepackage{subfigure}
\usepackage{subcaption}
\usepackage{tikz}
\usepackage{appendix}

\newcommand\mi{\mathrm{i}}
\newcommand\me{\mathrm{e}}

\newcommand\pp{\uppi}

\newcommand{\dif}{\mathrm{d}}

\begin{document}

\title{Parametric resonance amplification of gravitational waves in dynamical Chern-Simons gravity}

\author{Han-Wen Hu$^{1,2}$}
\email{huhanwen@itp.ac.cn}
\author{Chen Lan$^{3}$}
\email{stlanchen@126.com}
\author{Zong-Kuan Guo$^{1,2,4}$}
\email{guozk@itp.ac.cn}
	\affiliation{$^1$Institute of Theoretical Physics, Chinese Academy of Sciences, P.O. Box 2735, Beijing 100190, China}
	\affiliation{$^2$School of Physical Sciences, University of Chinese Academy of Sciences, No.19A Yuquan Road, Beijing 100049, China}
    \affiliation{$^3$Department of Physics, Yantai University, 30 Qingquan Road, Yantai 264005, China}
   \affiliation{$^4$School of Fundamental Physics and Mathematical Sciences, Hangzhou Institute for Advanced Study, University of Chinese Academy of Sciences, Hangzhou 310024, China}

\begin{abstract}
    Within the effective field theory of dynamical Chern-Simons (dCS) gravity, we study parametric resonance amplification of gravitational waves driven by an oscillating environmental field coupled to the dCS pseudoscalar.
    We find that the black hole potential barrier and external shell form a resonant cavity, producing a Mathieu instability whose optimal frequency is fixed by the cavity length.
    The instability shows a horizon leakage threshold, Floquet sidebands, and a delayed secondary burst in axial gravitational perturbations.
    This mechanism reveals that dCS corrections at ultraweak coupling can still accumulate via long term parametric amplification, leaving discernible signatures in gravitational wave signals.
\end{abstract}

\maketitle

\section{Introduction}

Gravitational-wave (GW) astronomy has moved tests of general relativity (GR) from weak field, quasi static systems such as the solar system and binary pulsars \cite{Will:2001mx,Bertotti:2003rm,Will:2014kxa} to the strong-field dynamics of compact binary mergers, where the LIGO-Virgo-KAGRA Collaboration has tested GR with hundreds of events \cite{LIGOScientific:2016dsl,DeLaurentis:2016jfs,LIGOScientific:2019fpa,LIGOScientific:2020aai,LIGOScientific:2021sio,Mehta:2022pcn,Payne:2024yhk,LIGOScientific:2025wao}.
The ringdown phase of binary black hole (BH) mergers is especially useful because it probes rapidly evolving high curvature spacetime \cite{Ghosh:2017gfp,LIGOScientific:2020tif,Krishnendu:2021fga} and can also carry imprints of environmental matter sources \cite{Yang:2026jch}.
From GW150914 to the latest statistical analyses of the GWTC-4 catalog, current data remain consistent with GR \cite{LIGOScientific:2016lio,LIGOScientific:2026qni,LIGOScientific:2026fcf,LIGOScientific:2026wpt}, while third generation detectors are expected to improve the ringdown signal-to-noise ratio by more than an order of magnitude \cite{Reitze:2019dyk}.
Since GR is nonrenormalizable and should be regarded as the low energy effective field theory (EFT) of quantum gravity, a central target of GW physics is to identify observable low energy corrections.
Dynamical Chern-Simons (dCS) gravity is a well-motivated example: it follows from high-energy consistency arguments rather than ad hoc phenomenology \cite{Jackiw:2003pm,Alexander:2007kv,Yunes:2009hc,Delsate:2014hba}, introduces a pseudoscalar field coupled to the Pontryagin density, and provides a minimal macroscopic test of parity violation in gravity \cite{Yagi:2012vf,Alexander:2009tp,Molina:2010fb,Hu:2025efp}.
Weak field tests, multimessenger neutron star observations, and present GW observations constrain the dCS coupling \cite{Nakamura:2018yaw,Silva:2020acr,Silva:2022srr}. 
The EFT treatment adopted here relies on a weak coupling expansion, for which the dimensionless parameter $\zeta\equiv\alpha/M^2$ is taken to satisfy $\zeta \ll 1$ \cite{Yunes:2009hc}.
In BH perturbation theory, the Schwarzschild background has no nontrivial pseudoscalar condensate and the coupling between gravitational perturbations and the dCS scalar enters only at $\mathcal{O}(\alpha)$, so the modified gravity signal is easily buried under standard quasinormal mode (QNM) decay \cite{Cardoso:2009pk,Molina:2010fb,Okounkova:2018pql,Wagle:2021tam}.
This motivates us to look beyond clean isolated BHs and study dirty BHs embedded in astrophysical environments \cite{Leung:1997was,Medved:2003pr,Barausse:2014tra,Bamber:2021knr,Maimon:2025hsi,Konoplya:2025ixm,Lan:2025brn}.

BHs in the universe are usually surrounded by relativistic plasma or cold dark matter halos \cite{EventHorizonTelescope:2021bee,Gondolo:1999ef,Macedo:2024qky,Kazempour:2024lcx,Ranjbar:2025yml},
and the dynamical evolution of surrounding matter can directly modify the BH gravitational waveform \cite{Tian:2025uvk}.
In standard GW environment models, two standard examples of these effects are
tiny long-term phase shifts from hydrodynamical processes \cite{Barausse:2014tra,Tomaselli:2023ysb,Mitra:2023sny,CanevaSantoro:2023aol}, and QNM instabilities from static gravitational potential corrections \cite{Cheung:2021bol,Konoplya:2022hll,Cardoso:2024mrw,Siqueira:2025lww,Oshita:2025ibu,Berti:2025hly}.
Related echo-like scattering effects can also produce spectral drift in ringdown signals \cite{Hu:2025beh}.
These passive effects are almost undetectable in current and future GW observations.
However, the role of the environment changes completely if it is not a static fluid, but a long lived local coherent real field background supported by new physics.
In static spacetimes, the dominant mode of such real fields oscillates at nearly a single frequency. 
It acts as a time dependent driving source that keeps feeding energy into the system, instead of just a static correction.
This time-dependent background field can drive the system and create the conditions for parametric resonance \cite{Fujita:2020iyx,Robbins:2021ucj,Chen:2024gqn,Kehagias:2025zws}.

Motivated by this possibility, we focus on whether a dynamical environment can make an otherwise weak dCS correction visible in ringdown signals.
Within the EFT of dCS gravity, we introduce a coupling between the environmental field $\chi$ and the dCS pseudoscalar as the leading low energy effective operator of the system \cite{Dvali:1995ce,Svrcek:2006yi,Marsh:2015xka,Burgess:2023ifd,Alexander:2024vav,Alexander:2025olg}.
Unlike the static environments usually considered before, here the environment is not a passive correction to the BH potential.
Instead, the Regge-Wheeler potential barrier and the outer oscillating shell form a resonant cavity, and the periodic environmental field acts as a pump for the dCS scalar.
When the driving frequency matches the cavity scale, the dCS scalar develops Floquet growth even for tiny modified-gravity coupling.
This amplified scalar then sources axial gravitational perturbations and leaves two observable signatures: delayed secondary bursts in the time domain and Floquet sidebands in the frequency domain.
The mechanism therefore gives a concrete route by which high-precision ringdown observations could probe weak dCS effects in dynamical astrophysical environments.

This paper is organized as follows. 
Sec.\ \ref{sec:II} builds the EFT for dCS gravity coupled to the environmental field and derives the linear perturbation equations.
Sec.\ \ref{sec:III} uses standard numerical simulations to study the parametric instability of dCS scalar modes.
Sec.\ \ref{sec:IV} presents the parametric resonance amplification mechanism of GWs, focusing on the Floquet sidebands of scalar perturbations and the secondary burst of gravitational perturbations in the time domain.
Sec.\ \ref{sec:V} gives a brief conclusion and discussion. 
We use geometric units with $c = G = 1$.

\section{Effective field theory framework and perturbation equations}\label{sec:II}

To study how the external environment affects perturbations of static spherically symmetric BHs in dCS gravity, we build an EFT model. 
It includes the standard dCS action and a local external real field $\chi$ with even parity.
$\chi$ does not have to be a fundamental particle. 
It represents the macroscopic bound state of the environment around the BH.
We ignore the dynamics of $\chi$ and only keep its coupling to the dCS scalar.
We adopt the following action
\begin{equation}\label{eq:full-action}
    S = \int \dif^4x \sqrt{-g} \left[R + \frac{\alpha}{4} \vartheta {}^*R R- \frac{\beta}{2}  g^{\mu\nu}\nabla_\mu \vartheta \nabla_\nu \vartheta + \mathcal{L}_{\rm int} \right],
\end{equation}
where $g \equiv |\det g_{\mu\nu}|$ is the determinant of the metric tensor and $R$ is the Ricci scalar.
The second term is the dCS modified gravity coupling, where $\vartheta$ is the dCS pseudoscalar, $\alpha$ is the dCS coupling constant, and ${}^*R R \equiv {}^{*}R^{\alpha \beta}_{\;\;\;\; \gamma \delta} R^{\gamma \delta}_{\;\;\;\; \alpha \beta}$ is the Pontryagin density \cite{Alexander:2009tp}.
The third term is the standard kinetic term of the dCS scalar, with $\beta$ a dimensionless constant.
Since the interaction term $\mathcal{L}_{\rm int}$ must be a Lorentz scalar and parity invariant, a leading order interaction reads
\begin{equation}
    \mathcal{L}_{\rm int} = - \lambda \chi \vartheta^2,
\end{equation}
where $\lambda$ is a coupling constant with dimension $[L]^{-2}$.
This coupling gives $\vartheta$ a spacetime dependent effective mass in the background $\chi(x)$
\begin{equation}
    m_{\rm eff,\vartheta}^2(x) = \frac{2\lambda}{\beta}\chi(x).
\end{equation}
Before proceeding, we comment on the naturalness of the coupling $\lambda$. 
As discussed in footnote \ref{fn:1stfn}, $\lambda$ inherits the UV scale $\Lambda_0^4$ from instanton dynamics. 
The effective driving amplitude is $\epsilon \equiv 2\lambda\chi_0/\beta = 0.1M^{-2}$, 
which with $\chi_0 \sim 10^{-3}$ gives $\lambda \sim 50\beta M^{-2}$. 
For a stellar-mass BH with $M \sim 10 M_\odot \sim 15\,\rm{km}$, 
this corresponds to $\lambda \sim 0.2\,\rm{km}^{-2}$. 
The associated UV scale is $\Lambda_0 \sim \lambda^{1/2} \sim 10^{-10}\,\rm{eV}$, 
safely below the dCS EFT cutoff $\Lambda_{\rm EFT} \sim \alpha^{-1/2}$. 
For $\alpha = 10^{-6}M^2$, $\Lambda_{\rm EFT} \sim 10^{-8}\,\rm{eV}$, 
so $\Lambda_0 < \Lambda_{\rm EFT}$ is satisfied. 
The fine-tuning required to maintain $\lambda$ at this scale is analogous to the hierarchy problem in spontaneous scalarization models \cite{Silva:2017uqg,Herdeiro:2018wub}, 
and we adopt the same pragmatic EFT stance: the operator is the lowest-order symmetry-breaking term, and its coefficient is treated as a free phenomenological parameter within the EFT validity window.

Notably, in a static background, $\chi$ is a real field, so its dominant time dependence takes the form $\chi \propto \cos(\Omega t)$.
Previous studies show such long-lived oscillations can exist for massive scalar fields outside Schwarzschild BHs \cite{Barranco:2011eyw,Barranco:2012qs}.
Thus, the environmental field introduces a localized, quasi-periodically oscillating mass into the dCS scalar equation
\footnote{
The dCS scalar $\vartheta$ has an approximate symmetry $\vartheta \to \vartheta + c$ in the massless limit.
The $\chi \vartheta^2$ term denotes the effective mass induced by the soft breaking of this symmetry.
Like axion fields, its shift symmetry is broken by nonperturbative effects such as instantons, generating a periodic potential $V(\vartheta) \sim \Lambda_{\rm UV}^4 [1-\cos(\vartheta/f)]$ \cite{Marsh:2015xka}.
If UV parameters setting the instanton strength (e.g., heavy fermion mass or gauge coupling) depend on the vacuum expectation value of $\chi$ \cite{Dvali:1995ce,Marsh:2015xka}, the characteristic scale becomes a function of $\chi$, $\Lambda_{\rm UV}(\chi) \simeq \Lambda_0 (1 + c_1 \chi/M_{\rm pl})$.
We expand $V(\vartheta, \chi)$. 
For $\chi \ll M_{\rm pl}$, the leading cross term yields $\mathcal{L}_{\rm int} \sim \Lambda_0^4 \chi \vartheta^2$.
This provides the microscopic origin of the interaction and shows that $\lambda$ inherits the ultraviolet scale $\Lambda_0^4$, explaining why $\lambda$ can be larger than the gravitational coupling. \label{fn:1stfn}
}.

To obtain the equations of motion (EoM), we vary the action \eqref{eq:full-action} with respect to the metric $g^{\mu\nu}$ \cite{Alexander:2009tp},
\begin{equation}\label{eq:EoM-gravity}
    G_{\mu\nu} + 2 \alpha C_{\mu\nu} = T_{\mu\nu}^{(\vartheta)} + T_{\mu\nu}^{\rm (int)},
\end{equation}
where $C_{\mu\nu}$ is the C-tensor, which includes the coupling of the gradient of $\vartheta$ with Riemann curvature.
It satisfies the traceless condition $g_{\mu\nu}C^{\mu\nu}=0$.
The source terms on the right-hand side are the energy momentum tensors from the $\vartheta$ and the interaction term.
We then vary the action with respect to the dCS scalar $\vartheta$,
\begin{equation}\label{eq:EoM-theta}
    \beta \Box \vartheta - 2 \lambda \chi \vartheta = - \frac{\alpha}{4} {}^*R R.
\end{equation}
Eq.\ \eqref{eq:EoM-theta} clarifies that when the environmental field $\chi$ oscillates quasi-periodically, the coefficients of the equation vary periodically in time, turning the scalar field equation into a Mathieu-type equation.
The resulting parametric instability is therefore a Mathieu instability, characterized by Floquet exponential growth inside the resonance bands.
This parametric driving allows the $\vartheta$ field to absorb energy from the environment and grow exponentially, even if the source term ${}^*R R$ is weak.
Eqs.\ \eqref{eq:EoM-gravity} and \eqref{eq:EoM-theta} form a coupled system. 
We solve it under specific physical assumptions.

We first consider the background solution.
For spherical symmetry, we have $C_{\mu\nu}={}^*R R=0$, so the right hand of Eq.\ \eqref{eq:EoM-theta} vanishes identically.
The interaction term is linear in $\vartheta$, so $\vartheta^{(0)} = 0$ is an exact solution, and no non-trivial dCS scalar condensate exists in the background.
The energy-momentum tensors from the dCS scalar and the interaction both vanish 
$T_{\mu\nu}^{(\vartheta)} = T_{\mu\nu}^{\rm (int)}=0$.
We further assume a weak external field, so the backreaction of $\chi$ on the background is negligible.
The background equation \eqref{eq:EoM-gravity} thus reduces to the vacuum Einstein equation, whose exact solution $g_{\mu\nu}^{(0)}$ is the Schwarzschild spacetime
\begin{equation}
    \dif s^2 = -f(r) \dif t^2 + f(r)^{-1} \dif r^2 + r^2 \dif \Omega^2, \quad f(r) = 1 - \frac{2 M}{r}.
\end{equation}
The weak-field condition $\chi \ll 1$ only suppresses the gravitational backreaction of the environment on the background, but does not limit the interaction strength $\mathcal{L}_{\rm int}$ for the dCS scalar.
In other words, the effective coupling strength $(2\lambda/\beta)\chi$ can drive strong parametric resonance if $\lambda/\beta$ is large enough.

We study a localized, single-peaked, time-oscillating matter shell outside the black hole. 
The simplest and most stable description is a finite-width spherically symmetric shell.
We thus model $\chi$ phenomenologically as a localized Gaussian oscillating shell
\begin{equation}\label{eq:background-chi-final}
    \chi^{(0)}(t,r)=\chi_0\exp\left[-\frac{(r-r_0)^2}{\sigma^2}\right]\cos(\Omega t).
\end{equation}
where $\chi_0$, $r_0$, $\sigma$ and $\Omega$ denote the background amplitude, the shell center, the effective width and the oscillation frequency of the dominant mode, respectively.
Astrophysically, $\chi$ can be interpreted as a coherent real scalar environment around the BH, such as a pulson/oscillon like scalar cloud or a compact dark sector configuration. 
A concrete field theoretic realization of the localized Gaussian profile is provided by the real scalar pulson solution with logarithmic self-interactions \cite{Koutvitsky:2005ds,Koutvitsky:2011yq}, in the present work we use its thin-shell, local-flat limit and keep only the leading cosine harmonic of the quasi-periodic motion, which gives Eq.\ \eqref{eq:background-chi-final}.

We now study linear perturbations of the metric and the dCS scalar.
We decompose the metric perturbation $h_{\mu\nu}$ into axial and polar parts on the spherical harmonic basis.
As shown in Ref.~\cite{Molina:2010fb}, the dCS pseudoscalar perturbation $\delta \vartheta$ only couples to axial perturbations.
In the Regge-Wheeler gauge, we expand the axial perturbation with tensor spherical harmonics and introduce the master variable $\Psi(t,r)$.
We also write the dCS scalar perturbation as $\delta \vartheta = r^{-1} \Theta(t,r) Y_{lm}$.
This gives a set of coupled linear perturbation equations \cite{Molina:2010fb,Hu:2025efp}:
\begin{subequations}\label{eq:pert-eq}
	\begin{equation}\label{eq:pert-eq-g}
		-\frac{\partial^2 \Psi}{\partial t^2} + \frac{\partial^2 \Psi}{\partial r_*^2} - f\left(\frac{l (l +1)}{r^2}-\frac{6 M}{r^3}\right) \Psi = \frac{6 M \alpha}{r^5} f \Theta,
	\end{equation}
	\begin{equation}\label{eq:pert-eq-s}
		-\frac{\partial^2 \Theta}{\partial t^2} + \frac{\partial^2 \Theta}{\partial r_*^2} - f\left[\frac{l (l +1)}{r^2}\left(1+\frac{36 M^2 \alpha^2}{r^6 \beta}\right)+\frac{2 M}{r^3} + \frac{2 \lambda}{\beta} \chi^{(0)}\right] \Theta = f \frac{(l +2)!}{(l -2)!} \frac{6 M \alpha}{r^5 \beta} \Psi.
	\end{equation}
\end{subequations}
Compared with standard dCS axial perturbation equations, Eq.\ \eqref{eq:pert-eq-g} remains unchanged.
This relies on the fact that the interaction energy momentum tensor $T_{\mu\nu}^{\rm (int)}$ is proportional to $\vartheta^2$, which is a second-order term.

In contrast, the dCS scalar equation \eqref{eq:pert-eq-s} is modified.
Compared with the vacuum case, its effective potential gains an extra term $(2\lambda/\beta)\chi^{(0)}$ induced by the environmental field.
From Eq.\ \eqref{eq:background-chi-final}, this term acts as a periodic pump. 
It triggers parametric resonance for the dCS scalar, which then draws energy from the background field and grows exponentially.

Current gravitational-wave observations agree well with general relativity, so the dCS coupling $\alpha$ is constrained to satisfy $\zeta \equiv \alpha/M^2 \ll 1$.
In this weak-coupling limit, the source terms in Eqs.\ \eqref{eq:pert-eq} are suppressed by $\alpha$, so gravitational and scalar perturbations nearly decouple.
Eq.\ \eqref{eq:pert-eq-g} reduces to the standard Regge-Wheeler equation, and the dCS scalar equation becomes the modified Klein-Gordon equation on Schwarzschild background.
However, although $\alpha$ is tiny, Eq.\ \eqref{eq:pert-eq-s} shows $\Theta$ grows exponentially driven by $\chi$.
As $\Theta$ is strongly amplified by parametric resonance, the source term $S_{\Psi} \propto \alpha \Theta$ can no longer be neglected.
Even with an extremely small $\alpha$, the resonantly amplified dCS scalar acts as a strong source and efficiently excites axial gravitational waves.
It makes tiny dCS effects potentially observable in ringdown signals.
In the next section, we numerically simulate this parametric resonance by time domain evolution of the coupled equations, to quantify the GW radiation features.

\section{Parametric resonance of scalar modes}\label{sec:III}

In the coupled perturbation equations \eqref{eq:pert-eq} derived above, the gravitational and dCS scalar sectors interact through the coupling $\alpha$.
In the weak coupling limit, the source term $S_{\Theta}$ on the right of Eq.\ \eqref{eq:pert-eq-s} describes the backreaction of GWs on the dCS scalar.
Since $S_{\Theta} \propto \alpha$, it is far weaker than the parametric driving term $2 \chi^{(0)} \Theta \lambda / \beta$.
This backreaction can thus be neglected at the early stage of the instability.

We consider the dCS scalar evolution on a fixed Schwarzschild geometry with a time-dependent environmental background.
The gravitational perturbation $\Psi$ becomes a secondary effect driven unidirectionally by $\Theta$.
In this section, we neglect $\mathcal{O}(\alpha)$ and $\mathcal{O}(\alpha^2)$ terms in Eq.\ \eqref{eq:pert-eq-s}. 
The master equation for the dCS scalar then reduces to the massless Klein-Gordon equation on Schwarzschild with a periodic bump,
\begin{equation}\label{eq:evolution-theta}
    -\frac{\partial^2 \Theta}{\partial t^2} + \frac{\partial^2 \Theta}{\partial r_*^2} - f(r) \left[ \frac{l(l+1)}{r^2} + \frac{2M}{r^3} + V_{\rm bump} \right] \Theta = 0,\quad V_{\rm bump}=\frac{2 \lambda}{\beta}\chi_0 \exp\left[-\frac{(r-r_0)^2}{\sigma^2}\right] \cos(\Omega t).
\end{equation}
Eq.\ \eqref{eq:evolution-theta} is the core of our numerical simulations.
By Floquet theory, perturbations grow exponentially when the real part of the intrinsic QNM frequency $\Re(\omega_n)$ and the external driving frequency $\Omega$ satisfy the parametric resonance condition.
We solve Eq.\ \eqref{eq:evolution-theta} using time-domain evolution to quantitatively verify the predicted Mathieu instability and determine its resonance bands.

To achieve numerical integration of Eq.\ \eqref{eq:evolution-theta}, we use the double null coordinate method widely adopted in dCS BH perturbation theory \cite{Wang:2000dt,Wang:2004bv,Molina:2010fb,Boudet:2022wmb}.
Setting $u = t - r_*$ and $v = t + r_*$, the evolution equation can be exactly rewritten in the first order form
\begin{equation}
    -4 \frac{\partial^2 \Theta}{\partial u \partial v} = V_{\rm eff}(u,v) \Theta,
\end{equation}
where the total effective potential reads $V_{\rm eff}(u,v) \equiv f(r) \left[ l(l+1)/r^2 + 2M/r^3 + V_{\rm bump}(u,v) \right] \equiv V_{l}(u,v)+ f(r) V_{\rm bump}(u,v)$.
This characteristic coordinate system naturally aligns physical boundaries with the grid. 
It automatically satisfies the pure ingoing boundary at the black hole horizon and pure outgoing boundary at infinity, eliminating artificial reflections from finite truncated boundaries.

We use the second order convergent finite difference scheme proposed by Gundlach, Price and Pullin \cite{Gundlach:1993tp,Konoplya:2011qq}.
On a two dimensional light cone grid with uniform step $h = \Delta u = \Delta v$, the field value at $N=(u+h, v+h)$  is explicitly determined by three adjacent points $S=(u, v)$, $W=(u+h, v)$ and $E=(u, v+h)$ on its past light cone
\begin{equation}
    \Theta_N = \Theta_W + \Theta_E - \Theta_S - \frac{h^2}{8} V_{\rm eff}(S) (\Theta_W + \Theta_E) + \mathcal{O}(h^4).
\end{equation}
Since $V_{\rm bump}$ contains the driving time $t = (v+u)/2$, we adopt the local coordinates at point $S$ when computing the effective potential.
We then specify initial data on the characteristic lines $u=u_0$ and $v=v_0$.
Considering realistic perturbations from binary mergers or infalling matter, we set a broad Gaussian wave packet propagating toward the horizon on the $v$-line, with zero initial perturbation at the horizon
\begin{equation}
    \Theta(u=u_0, v) = \exp\left[ -\frac{(v-v_c)^2}{2w^2} \right], \quad \Theta(u, v=v_0) = 0,
\end{equation}
where $v_c$ is the wave-packet center and $w$ is its effective width.
We set the black hole mass $M=1$ for dimensionless normalization and focus on quadrupole perturbations with $l=2$.

The key parameters in numerical simulations are the driving amplitude $\epsilon \equiv 2\lambda \chi_0/\beta$ and the frequency $\Omega$.
Based on our analysis of the EFT energy hierarchy, the large effective coupling ratio enables $\epsilon$ to be a freely adjustable phenomenological parameter within the range from $\mathcal{O}(0.1) M^{-2}$ to $\mathcal{O}(10) M^{-2}$.
For the simulations below, we set the effective amplitude to $\epsilon=0.1 M^{-2}$ and the Gaussian shell width to $\sigma=2M$.
Parameter scans over $(\Omega,\epsilon)$ show that increasing $\epsilon$ broadens the resonance band and shifts its central ridge toward larger $\Omega$.
Along the optimal ridge, the maximum Lyapunov exponent follows $\gamma_{\rm max}\propto \epsilon^{1.07}$ for $\epsilon M^2<1$, while strong driving gives a sublinear scaling $\gamma_{\rm max}\propto \epsilon^{0.62}$.
The value used below, $\epsilon=0.1M^{-2}$, therefore lies in the linear Mathieu-type regime.
Only the frequency $\Omega$ and the central position $r_0$ of the Gaussian shell remain tunable.
We first analyze the frequency response at fixed shell position and then scan the full $(r_0,\Omega)$ plane.

\subsection{Time domain waveform evolution signatures}

\begin{figure}[!htb]
    \centering
    \includegraphics[width=0.6\linewidth]{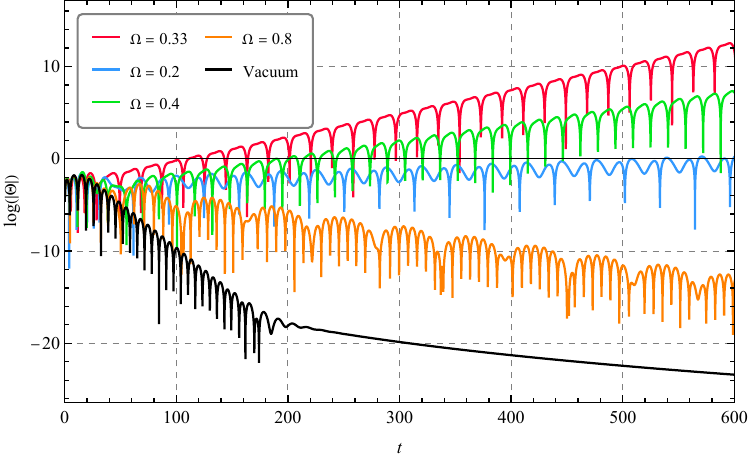}
    \caption{Logarithmic time-domain waveforms of the dCS scalar perturbation $|\Theta(t)|$ with $r_0=20M,\ \sigma=2M,\ \epsilon=0.1 M^{-2}$.
    The black solid line represents the pure Schwarzschild background, exhibiting standard QNM decay.
    For time dependent environmental fields with off resonance driving frequencies ($\Omega M=0.2,\ 0.4,\ 0.8$), interference between external driving and the intrinsic system frequency produces distinct persistent beating patterns in scalar perturbations.
    Some such parameters fall in the weak instability regime, so the late-time envelopes no longer decay as in the vacuum case, but grow slowly at a low gain rate.
    In contrast, when the driving frequency satisfies the resonant condition $\Omega M =0.33$, the scalar perturbation undergoes exponential growth after a brief transient scattering.}
    \label{fig:waveform}
\end{figure}

We fix $r_0=20 M$ and plot the logarithmic time evolution of the dCS scalar perturbation $|\Theta(t)|$ for different driving frequencies $\Omega$ in Fig.\ \ref{fig:waveform}.
The scalar evolution with no environmental field ($\epsilon=0$) is also presented for comparison.
The waveforms reveal that the scalar dynamics are sensitive to $\Omega$ and exhibit diverse evolutionary behaviors.
The scalar perturbation curve in the vacuum case exhibits typical ringdown decay dominated by QNMs.
Notably, the parametric resonance mechanism here differs greatly from conventional intuition.
If periodic external driving couples mainly to the intrinsic modes near the Regge-Wheeler potential, the strongest response is expected around the fundamental QNM frequency ($\omega_{\rm QNM} \simeq 0.48 M^{-1}$ for $l=2$), with the primary parametric resonance occurring at $\Omega \simeq 2\omega_{\rm QNM} \simeq 0.96 M^{-1}$.
However, Fig.\ \ref{fig:waveform} clearly shows the dominant resonance with the fastest exponential growth emerges at a much lower frequency $\Omega = 0.33 M^{-1}$.
This originates from the fact that the Regge-Wheeler barrier and the oscillating Gaussian potential form a resonant cavity.
The amplified modes are cavity modes trapped in this cavity, rather than the intrinsic BH QNMs.
We provide an analytical estimate as follows.
the effective cavity length in tortoise coordinates is $L_{\rm cav} \simeq 22.8 M$.
Approximating the cavity as a rigid wall potential well, the fundamental eigenfrequency reads $\omega_{\rm cav} \sim \pp / L_{\rm cav} \simeq 0.138 M^{-1}$.
The primary parametric resonance should appear near $\Omega \simeq 2 \omega_{\rm cav} \simeq 0.276 M^{-1}$.
Considering potential penetration and gravitational redshift near the BH, this analytical result agrees well with the driving frequency $\Omega = 0.33 M^{-1}$ from numerical simulations.

With the driving frequency detuned from the primary resonance ($\Omega M = 0.2,\ 0.4,\ 0.8$), the time domain waveforms show rich dynamics.
The frequency detuning causes phase interference between the driving field and the system response, generating distinct beating patterns. 
Unlike the pure decay in the vacuum case, high order sub-resonances allow the energy pumped by the environment to weakly overcome the intrinsic dissipation of the BH, leading to slow growth of the scalar field amid oscillations. 
When the driving frequency is tuned to the primary resonance window $\Omega M = 0.33$ (satisfying $\Omega \simeq 2 \omega_{\rm cav}$), the efficiency of parametric resonance reaches its maximum. 
The scalar perturbation rapidly transitions to exponential growth $\Theta(t) \propto \me^{\gamma t},\ \gamma > 0$.

\subsection{Excitation of cavity bound states and thresholds of waveform instability}

While the time-domain waveform for $r_0=20M$ intuitively shows the strong parametric resonance gain of cavity modes, single-point data cannot establish the scaling law $\Omega \propto 1/L_{\rm cav}$.
To verify that the growth originates from selective resonance for a specific cavity geometry, we extract global characteristic scalars from long-time evolution and perform a systematic scan in the two-dimensional parameter space $(r_0, \Omega)$ to plot phase diagrams. 
We scan $r_0$ from $10M$ to $30M$ with a step size $\Delta r_0 = M$, and scan the driving frequency $\Omega$ from $0.1M^{-1}$ to $M^{-1}$ with a step size $\Delta \Omega = 0.01M^{-1}$.

To quantify the perturbation growth rate, we introduce the asymptotic Lyapunov exponent $\gamma$,
\begin{equation}
\gamma(r_0, \Omega) = \lim_{t \to \infty} \frac{1}{t} \log |\Theta(t, r_{\rm obs})|.
\end{equation}
The system is defined as unstable if and only if $\gamma > 0$. 
In numerical calculations, we fit the linear slope of the envelope of $\log |\Theta(t,r_{\rm obs})|$ in a sufficiently late time window to extract this exponent.

\begin{figure}[!htb]
    \centering
    \subfigure[Heat map 
    of the asymptotic Lyapunov exponent $\gamma$ in the $(r_0, \Omega)$ plane. 
    The warm colored regions ($\gamma > 0$) clearly mark the Arnold tongue structures where parametric resonance instability occurs. 
    The optimal resonance frequency $\Omega_{\rm res}$ (black pentagrams in the plot) exhibits significant blue shift when $r_0$ decreases. ]
    {\includegraphics[width=0.4\linewidth]{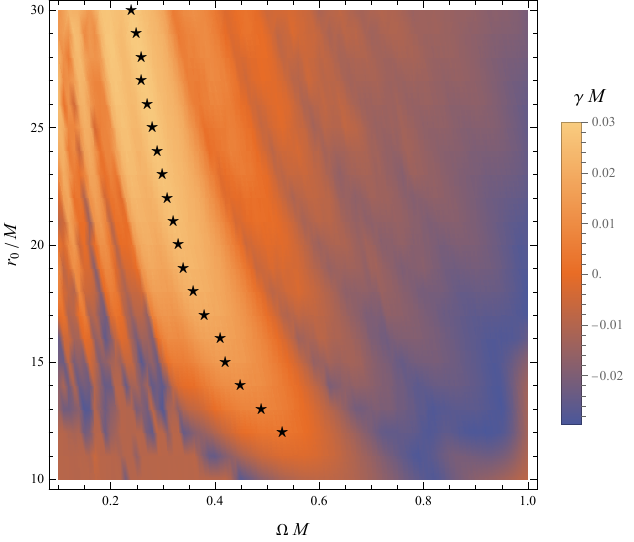}\label{fig:thermodynamic}}
    \subfigure[Evolution of the maximum Lyapunov exponent $\gamma_{\rm max}$ along the optimal slice (black pentagram in panel (a)) versus $r_0$. 
    For $r_0 > 12 M$, $\gamma > 0$, indicating stable exponential growth of scalar perturbations. 
    As the right side of the cavity approaches the BH, the quantum tunneling through the horizon barrier causes the growth rate to decrease sharply and cross the $\gamma=0$ threshold near $r_c \simeq 11.3 M$. 
    This drives the resonant cavity into an overdamped decaying state.]
    {\includegraphics[width=0.5\linewidth]{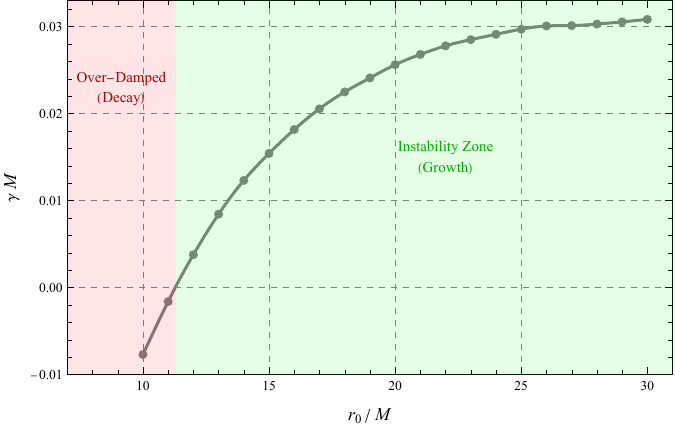}\label{fig:Lyapunov}}
    \caption{Two dimensional phase diagram of parametric resonance and the profile of the maximum Lyapunov exponent along the optimal frequency slice varying with $r_0$.}
    \label{fig:phase-diagram}
\end{figure}

Based on the scanning results, we construct the complete parametric resonance stability phase diagram shown in Fig.\ \ref{fig:phase-diagram} on the $(r_0, \Omega)$ plane and extract relevant dynamical quantities.
The numerical results firmly confirm the existence of unstable regions over a wide parameter range and reveal two universal physical laws governing this nonlinear dynamical system.

We first investigate the distribution of resonant bands in the phase diagram. 
If external driving mainly acts on the intrinsic potential barrier of the BH, the dominant response is expected at the inherent quasinormal mode $\omega M\simeq 0.48$. 
However, the heat map in Fig.\ \ref{fig:thermodynamic} clearly shows that resonant regions with notable positive growth rates ($\gamma > 0$) present tilted Arnold tongues in the parameter space.
The optimal driving frequency $\Omega_{\rm res}$ corresponding to the strongest instability (black pentagrams) is position dependent: 
$\Omega_{\rm res}$ undergoes blue shift as $r_0$ decreases. 
This frequency shift confirms our physical mechanism that the scalar perturbation growth arises from the resonance of bound states trapped in the resonant cavity.

\begin{figure}[!htb]
    \centering
    \includegraphics[width=0.5\linewidth]{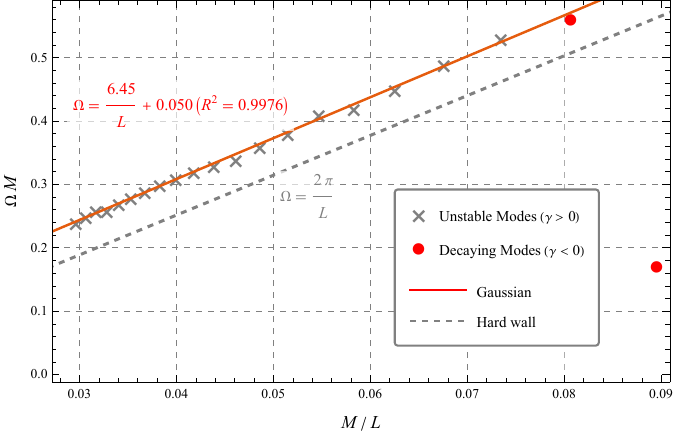}
    \caption{Quantitative verification of the resonant geometric law.
    The relation between the extracted optimal resonance frequency $\Omega_{\rm res}$ and the inverse cavity length $M/L$ is presented.
    The numerical data follow the linear fitting $\Omega M = 6.45(M/L) + 0.050$ ($R^2=0.9976$), almost consistent with the theoretical prediction $\Omega = 2\pp/L$ of ideal rigid cavity modes.
    This confirms the scalar instability essentially arises from selective resonance of bounded modes.}
    \label{fig:cavity}
\end{figure}

To quantitatively verify this picture, we define the effective characteristic cavity length as $L \simeq r_*(r_0) - r_*(3M)$.
For an ideal rigid wall potential well, the fundamental eigenfrequency satisfies $\omega_{\rm cav} = \pp / L$, and the corresponding primary parametric resonance frequency is $\Omega = 2\omega_{\rm cav} = 2\pp / L$.
We extract the optimal resonance frequency $\Omega_{\rm res}$ at each $r_0$ slice from the phase diagram and plot it as a function of the inverse cavity length $M/L$ in Fig.\ \ref{fig:cavity}; 
the numerical data exhibit an excellent linear scaling law.
Regression analysis yields the fitting relation
\begin{equation}
\Omega_{\rm res} M = \frac{6.45}{L/M} + 0.050, \quad (R^2 = 0.9976).
\end{equation}
The slope $6.45$ is consistent with the theoretical prediction $2\pp \simeq 6.28$ of an ideal rigid cavity.
Tiny deviations and the intercept arise from soft wall penetration of the Gaussian potential edge and gravitational redshift effects.
The high $R^2=0.9976$ confirms that the instability essentially originates from resonance with cavity eigenmodes at specific geometric scales.
The deviation of the fitted slope $6.45$ from the hard-wall prediction $2\pp \simeq 6.28$ and the nonzero intercept $0.050$ can be understood as follows. 
The Gaussian potential edge introduces an effective penetration depth $\delta r_* \simeq \sigma/2 = M$ beyond the nominal shell position, 
extending the cavity length by $\delta L \simeq M$. 
This yields a corrected eigenfrequency $\omega_{\rm cav}^{\rm soft} = \pp/(L + \delta L)$, shifting the resonance condition to
\begin{equation}
    \Omega_{\rm res} M = \frac{2\pp}{L/M + 1},
\end{equation}
which for $L/M \in [15, 35]$ predicts slopes and intercepts consistent with the fitted values to within $5\%$. 
The gravitational redshift near $r = 3M$ contributes an additional blueshift of order $(1 - 2M/r_{\rm cav})^{-1/2} - 1 \simeq 0.1$ to the effective frequency, accounting for the remaining discrepancy.

After establishing the geometric law for the optimal resonance frequency, we further explore how $\gamma$ varies with cavity length.
We extract the one-dimensional evolution of the maximum asymptotic Lyapunov exponent $\gamma_{\rm max}$ versus $r_0$ along the optimal resonance slice ($\Omega = \Omega_{\rm res}$), as shown in Fig.\ref{fig:Lyapunov}, and distinct dynamics emerge in the strong-field region.
For large $r_0$, the system stays in a stable parametric resonance regime with $\gamma_{\rm max} > 0$; 
$\gamma_{\rm max}$ rises with increasing $r_0$ and its growth gradually flattens out.
When the cavity is compressed to $r_0 \lesssim 18M$, the growth of $\gamma_{\rm max}$ slows and drops sharply near the critical radius $r_c \simeq 11.3 M$.
For all configurations with $r_0 < 11.3M$, the Lyapunov exponent turns negative, and the system evolves from parametric resonance into an overdamped decay state with $\gamma < 0$.
This means parametric resonance is suppressed for environmental fields which are too close to the BH, no matter how the driving frequency is fine-tuned.

To reveal the physical origin of the critical threshold $r_c \simeq 11.3M$, we establish the energy balance law for the dCS scalar field evolution.
In our EFT, the time-dependent background field $\chi(t,r)$ breaks the global time-translation symmetry, so the total energy of the scalar field is no longer conserved.
On the fixed Schwarzschild background metric $g_{\mu\nu}^{(0)}$, the action related to the dCS scalar field reads
\begin{equation}
S_{\vartheta} = \int \dif^4x \sqrt{-g}\ \left[ -\frac{\beta}{2} (\nabla \vartheta)^2 - \lambda \chi(t, r) \vartheta^2 \right].
\end{equation}
The energy-momentum tensor of $\vartheta$ is defined as
\begin{equation}
T_{\mu\nu}^{(\vartheta)} \equiv \beta \nabla_\mu \vartheta \nabla_\nu \vartheta - g_{\mu\nu} \left[ \frac{\beta}{2} (\nabla \vartheta)^2 + \lambda \chi \vartheta^2 \right],
\end{equation}
whose covariant divergence satisfies
\begin{equation}
\nabla^\mu T_{\mu\nu}^{(\vartheta)} = - \lambda \vartheta^2 \nabla_\nu \chi.
\end{equation}
The nonvanishing divergence indicates energy exchange is fully driven by the spacetime gradient of $\chi$.
We further define the energy flux vector $J^\mu_{\rm (E)} \equiv - T^\mu_{\;\;\nu} \xi^\nu = - T^\mu_{t}$ associated with the timelike Killing vector $\xi^\mu = (\partial_t)^\mu$ of the background geometry.
Using the Killing equation, the divergence of the energy flux yields
\begin{equation}
\nabla_\mu J^\mu_{\rm (E)} = - (\nabla_\mu T^\mu_{\;\;\nu}) \xi^\nu = \lambda \vartheta^2 \partial_t \chi.
\end{equation}
Integrating both sides over the spacelike hypersurface $\Sigma_t$ at fixed time $t$ (from the horizon $r_{\rm H}$ to spatial infinity) and applying the Gauss theorem, we obtain the macroscopic energy balance equation
\begin{equation}\label{eq:energy-balance}
\frac{\dif E_{\vartheta}}{\dif t} = P_{\rm pump}(t) + P_{\rm leak}(t).
\end{equation}
The parametric pumping power $P_{\rm pump}$ is defined by
\begin{equation}\label{eq:power-bump}
P_{\rm pump}(t) = - \lambda \Omega \int \dif^3x \sqrt{-g}\ \chi_0(r) \vartheta^2(r,t) \sin(\Omega t),
\end{equation}
which describes the power injected into the dCS scalar field by the $\chi$ background.\footnote{
If the environmental field $\chi$ is promoted to a responsive dynamical degree of freedom, the total energy of the $\chi$--$\vartheta$ subsystem is strictly conserved, and Eq.\ \eqref{eq:power-bump} exactly corresponds to the power pumped from the environmental field into the dCS scalar field.
In other words, the exponential growth of the dCS scalar field originates from the time-dependent pumping energy supplied by the external environmental field, instead of unphysical solutions that ought to be discarded.
}
In the parametric resonance regime, the dCS scalar field develops a stable phase difference $\delta$ with the time oscillating $\chi$ field, i.e., $\vartheta(t) \sim \me^{\gamma t} \cos(\Omega t / 2 + \delta)$.
Time averaging the power over one pumping period $T=2\pp/\Omega$ yields a net positive energy injection, $\langle P_{\rm pump} \rangle > 0$.
Since $\chi_0(r)$ is spatially distributed, the dependence of $P_{\rm pump}$ on $r_0$ is mainly governed by the spatial overlap between the cavity mode and the external shell.
Consequently, the variation of $P_{\rm pump}$ with $r_0$ is typically slow and non-exponential.
Furthermore, the horizon leakage power $P_{\rm leak}$ is given by the surface integral of the radial energy flux $J^r_{\rm (E)} = -T^r_{t}$ over the horizon area $\mathcal{A}$, and can be approximated as
\begin{equation}
    P_{\rm leak}(t) = - \int_{\rm H} J^r_{\rm (E)} \dif \mathcal{A} \simeq - \mathcal{C} |\vartheta(t, r_{\rm H})|^2 < 0,
\end{equation}
where $\mathcal{C}$ is a positive definite constant factor, and $P_{\rm leak}$ describes the power dissipated by the cavity mode into the BH horizon.
For the BH horizon with pure ingoing boundary conditions, this flux is always negative, representing energy loss.
For the dCS scalar mode $\vartheta$ trapped in the resonant cavity to reach the horizon, its wavefunction must penetrate the centrifugal barrier $V_l(r)$ of the BH.
When the energy level (i.e., the squared eigenfrequency $\omega_{\rm cav}^2 \simeq \Omega^2/4$) is lower than the barrier peak, this is a typical tunneling problem. 
Within the WKB approximation, the amplitude of the dCS scalar wavefunction at the horizon is exponentially suppressed by the potential barrier,
\begin{equation}
|\vartheta(r_{\rm H})|^2 \sim |\vartheta(r_0)|^2 \exp\left(- 2 \int_{-\infty}^{r_{\rm t}} \dif r_* \ \sqrt{V_{l}(r_*)-\omega_{\rm cav}^2}  \right),
\end{equation}
where $r_{\rm t}$ is the classical turning point.
This implies that the leakage power $|P_{\rm leak}|$ exhibits an extremely sensitive exponential dependence on the fundamental frequency of the resonant cavity.

This reveals the essence of the transition near the critical radius $r_c \simeq 11.3M$. 
As the environmental field approaches the BH, the shape of the Regge-Wheeler barrier remains largely unchanged, but the physical length $L$ of the resonant cavity is significantly compressed, inducing a strong blue shift of the intrinsic frequency of cavity modes. 
As the energy level $\omega_{\rm cav}^2$ is continuously raised to approach the peak of the barrier, both the integrand $\sqrt{V_{\rm eff} - \omega_{\rm cav}^2}$ and the width of the classically forbidden region shrink rapidly, leading to a rapid breakdown of the exponential suppression of $P_{\rm leak}$.
Further decreasing $r_0$ causes high-frequency cavity modes to directly transmit over the potential barrier. 
This process results in an exponential increase in the tunneling power to the horizon $|P_{\rm leak}|$, which surpasses the algebraically growing parametric pumping power. 
When $|P_{\rm leak}| > P_{\rm pump}$, the macroscopic energy derivative $\dif E_{\vartheta}/\dif t$ becomes negative, driving the system from parametric amplification into the overdamped decay regime. 
This indicates that in the model and parameter range considered here, when the external environment is too close to the horizon, the dissipative nature of the black hole dominates over the energy injection from parametric pumping; 
thus, only external environments sufficiently far from the horizon can act as a sustainable energy source for amplification.

\section{Parametric resonance amplification mechanism and gravitational wave observational signatures}\label{sec:IV}

We have established the existence and threshold of scalar Mathieu instability inside the GW resonant cavity composed of the Regge-Wheeler potential barrier and the external environment.
Nevertheless, the scalar sector in dCS theory cannot be directly measured by GW interferometers.
To convert the scalar modes into detectable physical signals, we explore their unique frequency domain features and investigate how the exponential growth of scalar perturbations is transferred into the GW sector via interactions.
In this section, we uncover the potential observational imprints of this mechanism on gravitational radiation from two perspectives: 
spectral fingerprint extraction and time domain amplification.

\subsection{Spectral fingerprints of scalar perturbations}

To identify observational features distinguishing ordinary forced oscillations from environmental parametric resonance, we perform Fourier transforms on the dCS scalar waveforms $\Theta(t)$ in the steady growth stage and investigate the spectral structure $\Theta(\omega)$ in the frequency domain.
In forced oscillation models driving at frequency $\Omega$, the dominant spectral peak of the long-term response is generally locked at $\omega = \Omega$.
Nevertheless, the scalar field governed by Eq.\ \eqref{eq:evolution-theta} features an intrinsic Mathieu-type parametric driving structure, where the system is driven by periodic modulation of equation parameters rather than direct external forcing.
For fundamental parametric resonance, the system absorbs external energy most efficiently when its response frequency is locked near half of the driving frequency.
To verify this mechanism across the parameter space, we select all growth modes ($\gamma > 0$) from the previous two-dimensional scan, extract their dominant response frequencies $\omega_{\rm peak}$, and plot the scatter distribution of $\omega_{\rm peak}$ versus the driving frequency $\Omega$.

\begin{figure}[!htb]
    \centering
    \subfigure[The dominant response frequency $\omega$ extracted from time-domain waveforms with $\gamma > 0$ in the full parameter space versus the external driving frequency $\Omega$.
    Colors denote different central positions $r_0$.
    All numerical data lie on the line $\omega = \Omega/2$, ruling out the ordinary forced oscillation mechanism with $\omega = \Omega$.]{\includegraphics[width=0.45\linewidth]{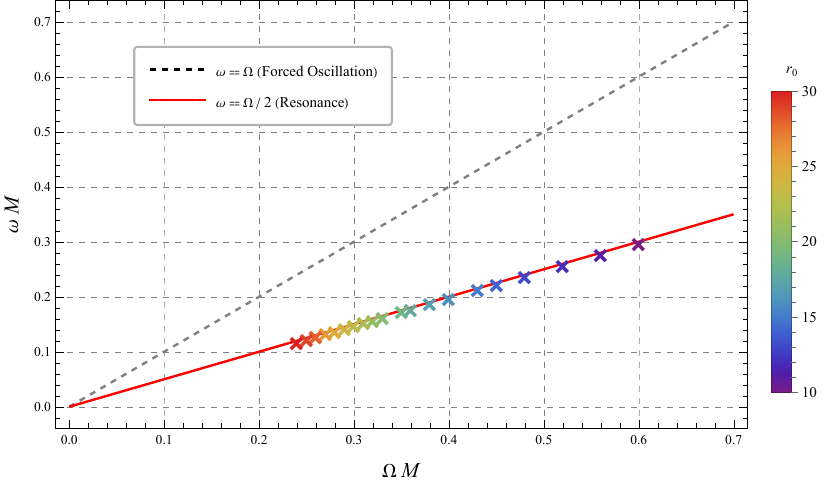}\label{fig:verify-resonance}}
    \subfigure[Logarithmic power spectral density of the scalar field in the late exponential-growth stage for $r_0=10M,\ 20M,\ 30M$.
    Apart from the main peak at $\omega = \Omega/2$, the spectrum clearly exhibits the first order Floquet sideband at $\omega = 1.5\Omega$ ($\omega + \Omega$).]{\includegraphics[width=0.45\linewidth]{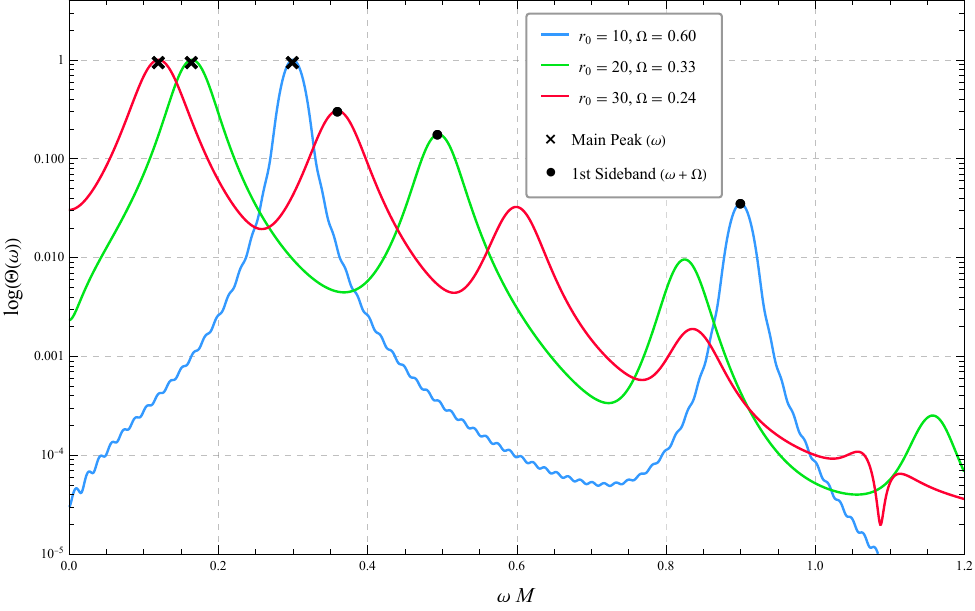}\label{fig:sideband}}
    \caption{Parametric resonance fingerprints and Floquet sidebands in the spectra of scalar perturbations}
    \label{fig:spectral-fingerprint}
\end{figure}

As shown in Fig.\ref{fig:verify-resonance}, for all growth modes under different spatial configurations, the numerically extracted peak frequency $\omega_{\rm peak}$ is distributed along $\omega = \Omega/2$, rather than the forced oscillation line $\omega = \Omega$.
This provides direct frequency domain evidence for parametric resonance. 
It demonstrates that the dCS scalar field does not follow the external oscillation via ordinary forced vibration, but behaves as a cavity bound state and continuously absorbs energy from the environmental field under resonant conditions.

Deeper spectral features manifest in the discrete harmonic structure induced by parametric driving.
According to Floquet theory, the solution of a linear system with periodic coefficients of frequency $\Omega$ can be expanded as a superposition of frequency components shifted by the driving frequency
\begin{equation}
\Theta(t) = \me^{\gamma t} \sum_{k=-\infty}^{\infty} c_k \me^{-\mi (\omega_{\rm peak} + k \Omega) t},
\end{equation}
where $\gamma$ is the Lyapunov exponent discussed previously, and $c_k$ represent the complex amplitudes of each harmonic component.
Substituting $\omega_{\rm peak} = \Omega/2$ into the above formula, we theoretically predict that besides the dominant resonant peak at $\Omega/2$ ($k=0$), a series of high-order sidebands spaced by $\Omega$ will arise in the spectrum, with peak frequencies at $1.5\Omega,\ 2.5\Omega,\ \cdots$.

To verify this theoretical prediction, we present the logarithmic power spectra of the scalar field in the exponential growth regime for three typical configurations $r_0 = 10M,\ 20M,\ 30M$ in Fig.\ \ref{fig:sideband}.
The numerical results agree well with Floquet theory.
Taking $r_0=20M$ (red curve) as an example, the optimal driving frequency satisfies $\Omega M=0.33$.
In addition to the dominant peak at $\omega M = 0.165$, the first order Floquet sideband is detected at $\omega M \simeq 0.495$ on the high-frequency side.
Similarly, for $r_0=30M$ (blue curve), the dominant peak and its corresponding first sideband also follow the relation $\Omega/2$ and $1.5\Omega$.
Moreover, the relative intensity ratio between the dominant peak and the sideband, namely $c_1/c_0$, varies with the system parameters.

This spectral fingerprint consisting of the dominant resonant peak and high order Floquet sidebands possesses distinct observational significance.
In modified gravity theories under static backgrounds, corrections to the effective potential from BH scalar hair or environmental effects generally only induce shifts of the main frequency, without naturally producing discrete sideband structures generated by dynamic parametric pumping.
For the dynamic external environment investigated in this work, the GW source term can inherit the discrete spectral ladder originating from parametric resonance.

\subsection{Delayed secondary burst of the gravitational perturbation}

After presenting the frequency domain fingerprint of parametric resonance for the dCS scalar field, we address a more crucial question: 
how does the dynamics in the scalar sector convert into detectable signals?
Current GW observations are highly consistent with the predictions of GR.
This strongly implies physically that if dCS modified gravity effects exist, the dimensionless coupling constant $\zeta \equiv \alpha/M^2$ must be extremely small.
In conventional perturbative analysis, such weak coupling only introduces nearly unobservable tiny corrections to the axial gravitational perturbation waveform $\Psi$.
To demonstrate how this mechanism changes the conventional understanding within full-coupling evolution, we perform numerical simulations with a representative small parameter $\alpha = 10^{-6} M^2$.

\begin{figure}[!htb]
    \centering
    \subfigure[Evolution of the fully coupled dCS scalar field, which agrees well with the approximation neglecting gravitational backreaction shown in Fig.\ref{fig:waveform}. 
    This indicates that the scalar dynamics are dominantly governed by the external environmental driving within the considered parameter range.]{\includegraphics[width=0.45\linewidth]{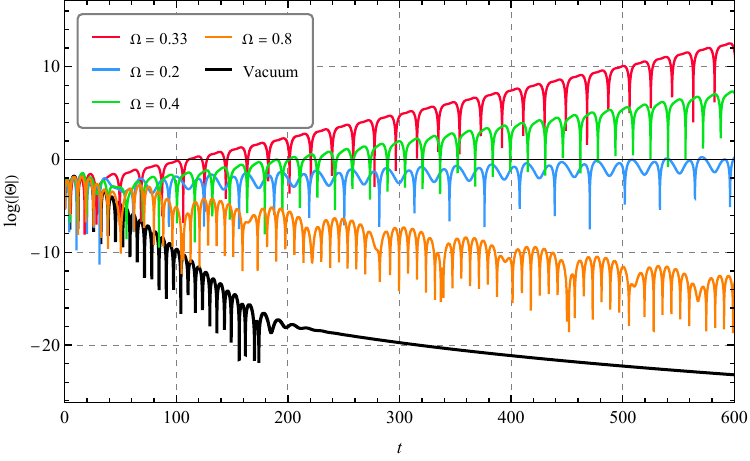}\label{fig:waveform-scalar}}
    \subfigure[Evolution of axial gravitational perturbations. 
    A prominent secondary burst emerges after the initial decay under resonant driving.]{\includegraphics[width=0.45\linewidth]{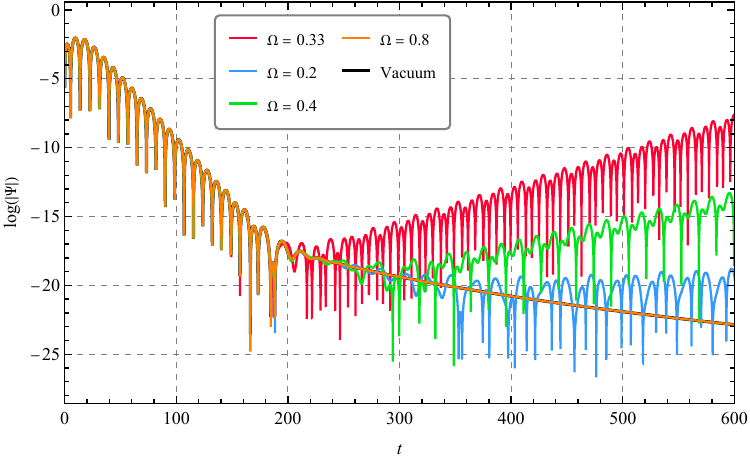}\label{fig:waveform-gravity}}
    \subfigure[Resonant case, $\Omega M=0.33$]{\includegraphics[width=0.32\linewidth]{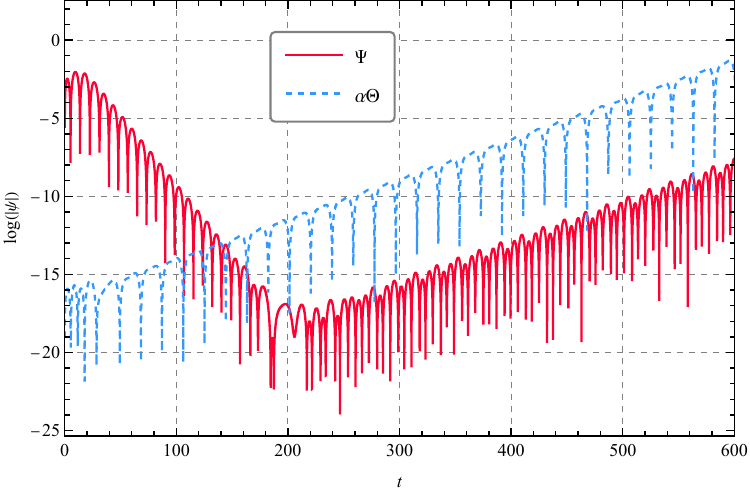}\label{fig:waveform-resonance}}
    \subfigure[Detuned case, $\Omega M=0.2$]{\includegraphics[width=0.32\linewidth]{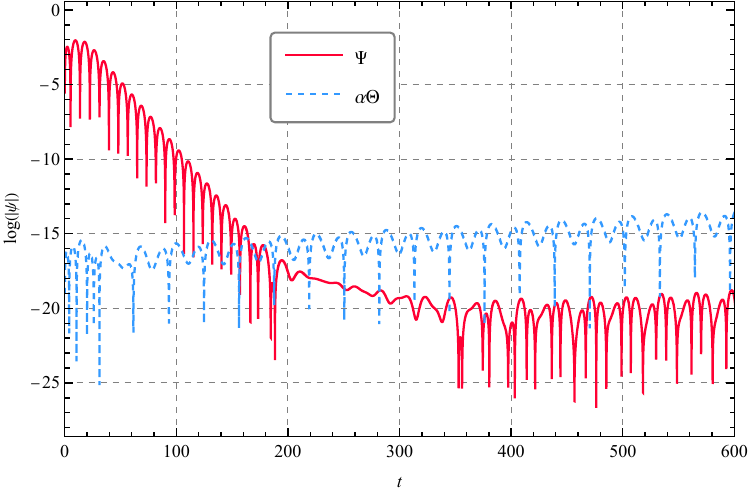}\label{fig:waveform-dissonance-1}}
    \subfigure[Detuned case, $\Omega M=0.8$]{\includegraphics[width=0.32\linewidth]{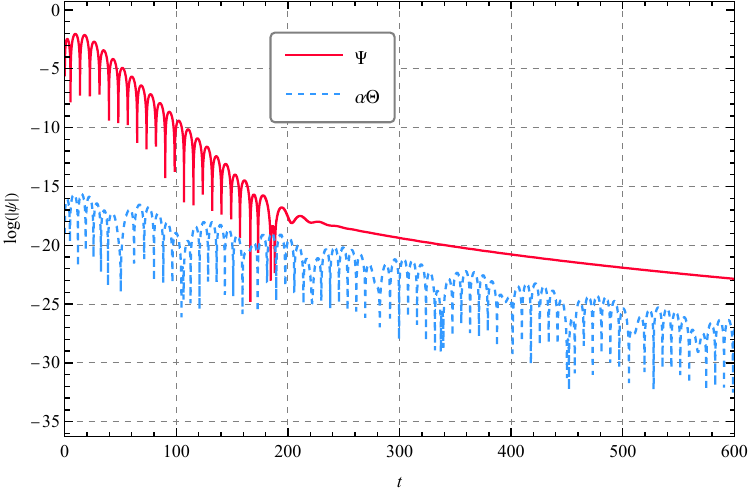}\label{fig:waveform-dissonance-2}}
    \caption{Panels (a) and (b) present the dynamical evolutions governed by the fully coupled perturbation equation \eqref{eq:pert-eq} for three scenarios: 
    vacuum, detuned driving, and fundamental parametric resonance. 
    Panels (c)-(e) compare the gravitational-wave amplitude $|\Psi|$ and the effective source term $\alpha|\Theta|$.
    We find that for fundamental resonance and mild detuning, the growing effective source term $\alpha|\Theta|$ (blue dashed line) overtakes the initially decaying gravitational-wave amplitude (red solid line). 
    Soon afterward, the axial perturbation equation is dominated by the source term $S_{\Theta}$, and the gravitational perturbation waveform departs from the pure dissipative ringdown of standard general relativity and enters a forced response regime. 
    Such a crossover does not occur in the strongly detuned case.
    All plots adopt $\alpha=10^{-6}$ and $r_0=20M$.}
    \label{fig:secondary-burst}
\end{figure}

We revisit the fully coupled dCS perturbation equation \eqref{eq:pert-eq} with a periodically varying source term.
For simplicity, we fix the center of the external environmental shell at $r_0 = 20M$.
We adopt the same driving parameters as in Fig.\ \ref{fig:waveform}, i.e., the resonant driving frequency $\Omega M = 0.33$ satisfying the fundamental resonance condition, together with the detuned frequencies $\Omega M = 0.2,\, 0.4,\, 0.8$.
For comparison, the perturbation profile in the vacuum case is also plotted.
The time-domain evolutions of the dCS scalar field $\Theta$ and the axial gravitational perturbation $\Psi$ are displayed in Fig.\ \ref{fig:secondary-burst}.

First, we focus on the backreaction of axial gravitational perturbations on the scalar field dynamics.
By comparing the scalar perturbation evolution curves in Fig.\ \ref{fig:waveform-scalar} and Fig.\ref{fig:waveform}, one can clearly see that although gravitational backreaction was neglected when we studied the parametric resonance characteristics between the external environment and the dCS scalar field in Sec.\ \ref{sec:III}, the numerical results are highly consistent with those from solving the fully coupled equations with inhomogeneous source terms here.
This numerical evidence supports our fundamental conclusion within the perturbative framework. Because the coupling constant $\alpha$ is extremely tiny, the backreaction of gravitational perturbations on the scalar field is almost negligible in the parametric resonance region we explore.

We now turn to analyze gravitational perturbations, where Fig.\ \ref{fig:waveform-gravity} reveals the nontrivial dynamical features of GW signals over time.
In the early stage as the initial wave packet impinges on the Regge-Wheeler potential barrier, the gravitational perturbation waveforms for all driving frequencies $\Omega$ exhibit nearly identical ringdown behaviors to those of a Schwarzschild BH.
We perform damped oscillation fitting on this early waveform segment to extract the quasinormal mode frequencies $\omega_i$, expressed as
\begin{equation}
\Psi(t)=\sum_{i=0}^n A_i \me^{\Im(\omega_{i})t} \cos \left[\Re(\omega_i)t + \phi_i \right],
\end{equation}
where $n$ is the number of extracted modes.
Using the standard Prony method, we find that all gravitational perturbation waveforms in Fig.\ \ref{fig:waveform-gravity} are dominated by the fundamental mode and first overtone of the standard Schwarzschild BH for $t< 180 M$,
\begin{equation}
M \omega_0 = 0.374-0.0889 \mi, \quad M\omega_1 = 0.347-0.178 \mi.
\end{equation}
These results demonstrate that within the time window $t< 180 M$, the tiny coupling constant $\alpha=10^{-6} M^2$ hides the dCS modified gravity effects behind the predictions of GR.
However, after the initial exponential decay, for the resonant driving frequency $\Omega M=0.33$ or mildly detuned cases with $\Omega M=0.2$, the axial gravitational perturbation amplitude shows a prominent and distinguishable secondary burst, and subsequently enters a forced oscillation stage with a growth rate comparable to that of scalar perturbations.

To reveal the dynamical mechanism behind the secondary burst, we compare the gravitational perturbation amplitude $|\Psi|$ and the effective source term $\alpha|\Theta|$ in Figs.\ \ref{fig:waveform-resonance}, \ref{fig:waveform-dissonance-1}, and \ref{fig:waveform-dissonance-2}.
Mathematically, the gravitational perturbation governed by Eq.\ \eqref{eq:pert-eq-g} is a linear superposition of the homogeneous intrinsic dissipative solution and the inhomogeneous particular solution.
At early evolutionary stages, due to the extremely small $\alpha$, the initial amplitude of the source term is submerged in the decaying background, satisfying $\alpha|\Theta| \ll |\Psi|$, so the dynamics of the gravitational field equation are dominated by the QNM solution ($\Psi \propto \me^{-|\Im(\omega)| t}$).
However, owing to the introduced time dependent dynamical external background, when the driving frequency $\Omega$ is tuned to resonance or slightly detuned, the source term grows approximately exponentially under sustained parametric pumping, i.e., $\alpha|\Theta| \propto \me^{\gamma t}$.
These two components with opposite evolutionary trends intersect at a critical time in the time domain.
We further provide a semi-quantitative analytical estimation for this process.
The decay of the homogeneous solution is dominated by the fundamental quasinormal mode $\omega_0$.
By equating the amplitude of the homogeneous solution $|\Psi(t)| \sim \me^{-|\Im(\omega_0)| t}$ with the growing amplitude of the inhomogeneous source term $\alpha|\Theta(t)| \simeq \alpha \me^{\gamma t}$, we obtain the critical crossing time $t_{\rm cross}$ when the inhomogeneous source begins to dominate the gravitational perturbation equation
\begin{equation}
t_{\rm cross} \simeq - \frac{1}{\gamma + |\mathrm{Im}(\omega_0)|} \log \alpha +\Delta t,
\end{equation}
where $\Delta t > 0$ accounts for the time delay between the amplitude crossing instant and the actual secondary burst occurrence.
This relation indicates that although the tiny coupling $\alpha \ll 1$ significantly postpones the secondary burst, the Lyapunov exponent $\gamma$ originating from parametric resonance in the denominator effectively reduces $t_{\rm cross}$.\footnote{
In realistic astrophysical merger events, the initial excitation amplitude of the scalar field is suppressed by the coupling constant as $A_{\Theta}/A_{\Psi} \propto \alpha$, such that the logarithmic term on the right-hand side becomes $-2 \log \alpha$, further delaying the burst time.
}

Furthermore, as shown in Figs.\ \ref{fig:waveform-resonance} and \ref{fig:waveform-dissonance-1}, shortly after the exponentially growing effective source term $\alpha|\Theta|$ overtakes the decaying homogeneous gravitational-wave amplitude, the gravitational perturbation equation becomes dominated by the inhomogeneous source term.
Thereafter, the gravitational waveform deviates from the ringdown decay; 
its late-time oscillation follows the growth trend of the scalar source and enters a stage governed by forced response.
The essential difference among various driving frequencies lies in the onset time at which the source term dominates Eq.\ \eqref{eq:pert-eq-g}.
For the fundamental resonance case with $\Omega M=0.33$ (Fig.\ \ref{fig:waveform-resonance}), the scalar field possesses a large Lyapunov exponent $\gamma$, and the scalar source $\alpha\Theta$ rapidly exceeds the perturbative magnitude, triggering a prominent secondary burst around $t \simeq 220M$.
For the mildly detuned case (Fig.\ \ref{fig:waveform-dissonance-1}), although the system cannot excite the strongest fundamental cavity mode resonance, it still falls into a weaker unstable band and maintains slow positive energy gain.
Consequently, the secondary GW burst is significantly postponed and gradually emerges in the late power law tail regime.
In the strongly detuned case with $\Omega M=0.8$ (Fig.\ \ref{fig:waveform-dissonance-2}), parametric resonance cannot overcome the dissipation induced by the BH horizon.
Under such parameters, the effective source term $\alpha|\Theta|$ undergoes overall decay and remains lower than the black-hole ringdown background throughout the entire simulation window, i.e., $\alpha|\Theta| \lesssim |\Psi|$ always holds.
In this scenario, the macroscopic GW signal reduces to the standard QNM ringdown of a vacuum Schwarzschild BH, and the extremely weak dCS effects are concealed within the standard ringdown signal.

We term this process the parametric resonance amplification of GWs: the environmental potential supplies parametric pumping that triggers exponential growth of the dCS scalar field, allowing the effective source term to exceed the perturbative magnitude under tiny coupling and ultimately inducing delayed secondary bursts of GWs.
This mechanism indicates that an extremely weak dCS coupling is not always unobservable in gravitational radiation.
In our model, as long as dynamical environments surrounding compact objects satisfy specific driving conditions, extremely weak modified gravity degrees of freedom can be accumulated through long-term parametric amplification and elevated to macroscopically detectable scales.

It is noteworthy that the exponential growth obtained within the linear perturbative framework cannot sustain indefinitely in realistic situations.
In a fully self-consistent model, when the resonantly amplified dCS scalar field builds up an energy density comparable to that of the external environmental background, its backreaction will gradually deplete the external energy reservoir.
Straightforward energy conservation estimates show that the depletion timescale $T_{\rm depl}$ of a typical external energy reservoir is much longer than the typical ringdown observation window $T_{\rm obs} \sim 10^2 M$.
Hence, approximating GWs as being driven by a nearly constant external energy source is physically valid for the evolutionary period of interest.

The possible observational relevance of this waveform response depends on the mass scale of the BH and on the coherence of the surrounding scalar environment. 
For the representative resonant case $\Omega M=0.33$, the dominant response frequency and the first Floquet sideband are located at $\omega M \simeq 0.165$ and $\omega M \simeq 0.495$, respectively. 
In physical units, this gives
\begin{equation}
f_{\rm peak}\simeq 180 {\rm Hz}\left(\frac{30M_\odot}{M}\right),\quad
f_{\rm sb}\simeq 530{\rm Hz}\left(\frac{30M_\odot}{M}\right).
\end{equation}
Thus, for stellar mass BHs with $M \sim 10$-$100 M_\odot$, the predicted spectral structures can lie in the frequency range of ground-based detectors. 
For massive BHs with $M \sim 10^6 M_\odot$, the same scaling shifts the signal to the mHz band, relevant for space based detectors. 
The effect is expected to be most visible when the external scalar environment is long-lived, coherent, and located outside the near-horizon leakage-dominated region, so that it can sustain the cavity resonance over the ringdown time scale.
A full detector level forecast is beyond the scope of the present work; here we identify the source-side waveform fingerprints, namely the Floquet sidebands and the delayed secondary burst, that future data analysis studies can search for.

\section{Conclusion and Discussion}\label{sec:V}

Starting from the dCS effective action supplemented by a local $\chi\vartheta^2$ operator, we followed the perturbation dynamics from the scalar channel to the axial gravitational waveform.
The external matter sector was treated as a coherent finite-width oscillator, which gives the dCS scalar a periodically varying effective mass on the Schwarzschild background.
Analytical estimates and time-domain evolutions show that the relevant frequency is selected by the trapped scalar mode between the inner gravitational potential barrier and the outer matter profile.
Near $\Omega\simeq2\omega_{\rm cav}$, the injected environmental energy overcomes the leakage through the horizon and drives exponential scalar growth.
Through the dCS mixing term, this growing scalar component is then converted into a late-time gravitational response, with a secondary wave packet and a discrete spectral ladder separated by the pump frequency.

The central finding is a matter-powered instability of the dCS degree of freedom whose scale is fixed by the exterior geometry rather than by the vacuum QNM spectrum.
This separates it from axion-background amplification of stochastic GWs \cite{Fujita:2020iyx,Peng:2022ttg} and from scalar-cloud growth associated with superradiance or stimulated decay \cite{Press:1972zz,Boskovic:2018lkj,Brito:2015oca}.
It also differs from the minimally coupled scalar setup of Ref.~\cite{Kehagias:2025zws}, where the resonance is tied to QNMs; in the present model the active frequency satisfies $\omega_{\rm cav}\ll\omega_{\rm QNM}$, and the amplified scalar is passed to the axial gravitational sector only through the dCS interaction.
The flux-balance analysis adds a spatial selection rule: matter configurations too close to the horizon leak energy faster than they pump it into the scalar mode, whereas sufficiently distant structures can sustain the instability over the ringdown window.
Thus the signal is not only a test of an ultraweak modified-gravity coupling, but also a probe of where coherent matter distributions, such as dark-matter spikes around central BHs, can reside \cite{Gondolo:1999ef,Karydas:2024fcn,Kavanagh:2024lgq}.

To estimate how spin modifies our results, 
we employ the slow-rotation approximation. 
For a Kerr BH with dimensionless spin $a_* = a/M \ll 1$, 
the Regge-Wheeler potential acquires a spin-dependent correction $\delta V_l \simeq -4ma_*M\omega/r^3$, 
where $m$ is the azimuthal quantum number and $\omega$ the mode frequency. 
For $l=m=2$ and $\omega \simeq \omega_{\rm cav} \simeq 0.14M^{-1}$, 
this shifts the effective cavity eigenfrequency by $\delta\omega_{\rm cav}/\omega_{\rm cav} \simeq 2a_* M^2/r_0^3 \times (4m\omega_{\rm cav}/\omega_{\rm cav}^2) \sim 0.1 a_*$ for $r_0 = 20M$. 
Thus for moderate spins $a_* \lesssim 0.3$, the resonance condition $\Omega \simeq 2\omega_{\rm cav}$ shifts by less than 10\%, 
and the cavity resonance mechanism remains operative. The full Kerr analysis is deferred to future work.

Several EFT caveats should be kept explicit.
The cubic interaction used in this work breaks the approximate shift symmetry of the dCS pseudoscalar and can generate quadratically divergent self energy corrections, so its coefficient should be treated as a phenomenological low energy parameter with an associated fine tuning cost.
Such lowest order shift symmetry breaking operators are also used in spontaneous scalarization models, where curvature couplings give scalar fields negative effective masses and trigger compact object instabilities \cite{Silva:2017uqg,Herdeiro:2018wub}.
As a check beyond the cubic model, we also tested the four-point interaction $\mathcal{L}_{\rm int}^{(4)}=-\lambda_4\chi^2\vartheta^2$.
Since $(\chi^{(0)})^2\propto [1+\cos(2\Omega t)]/2$, the effective scalar mass contains a static shift and an oscillating part at frequency $2\Omega$.
The resulting instability is much weaker: most channels remain damped, and the largest finite-time exponent is only $\gamma_{\rm max}M\simeq1.98\times10^{-3}$ near $r_0/M=29$ and $\Omega M=0.171$.
The least damped or weakly growing ridge still follows the effective cavity length and corresponds to the second cavity harmonic, supporting the geometric mode selection picture while showing that broad strong amplification is not generic for higher order couplings.

In this work, we worked on a static spherically symmetric Schwarzschild background, which is an exact solution of dCS gravity and therefore isolates the basic amplification mechanism without additional geometric assumptions.
Real compact objects, however, generally rotate, so the next step is to extend the model to Kerr spacetime.
In that case scalar clouds formed by superradiant instability may provide the external oscillating background dynamically, rather than being imposed as a external matter shell.
This makes non-spherically symmetric resonance in Kerr, nonlinear saturation under strong late-time backreaction, and derivative or higher order couplings that preserve shift symmetry natural directions for future work.
These extensions will decide whether the environment induced instability discussed here can be embedded in a more self consistent low energy description.

\vspace{5pt}
\noindent
\section*{Acknowledgments}

This work is supported by the National Natural Science Foundation of China No.~12475067 and No.~12235019.
Moreover, C. Lan is supported by Yantai University under Grant No.~WL22B224.

\bibliographystyle{apsrev4-1}
\bibliography{main}

\end{document}